\documentclass[a4paper,11pt]{article}
\pdfoutput=1 

\usepackage{jheppub} 
 \usepackage {caption}

\usepackage[T1]{fontenc} 

\newcommand{\ra}{\rightarrow}

\newcommand{\dd}{\mathrm{d}}

\newcommand{\be}{\begin{equation}}
\newcommand{\ee}{\end{equation}}

\title{Thermo-electric Transport of Dyonic Gubser-Rocha Black Holes}

\author[a]{Xian-Hui Ge,}

\author[a,1]{Zhaojie Xu\note{Corresponding author.}}

\affiliation[a]{Department of Physics, Shanghai University, \\Shanghai 200444, P.R. China}

\emailAdd{ge$\textunderscore$xh@hotmail.com}
\emailAdd{zeezj@shu.edu.cn}

\abstract{We study the thermo-electric transport coefficients of an extended version of the Gubser-Rocha model. After reviewing the two relaxation time model from holography and studying the effect of the magnetic field on thermo-electric transports from hydrodynamic theory, we present a new dilatonic dyonic asymptotically AdS black hole solution. Notice that S-duality plays an important role in finding the analytic solution with the magnetic field. Using the AdS/CMT dictionary, we analyze the electric and thermo-electric transport properties of the dual field theory. The resistivity exhibits T-linearity in the low-temperature regime. However, in the strong momentum relaxation and a strong magnetic field limit, the resistivitiy shows explicit deviation from the linear-in-T resistivity.  The Hall angle is linear in T for both the low-temperature regime and the high-temperature regime for fixed momentum dissipation strength. The Nernst signal is a bell-shaped function in terms of the magnetic field even when the momentum relaxation is strong. Finally, we discuss the possibility of getting a semi-realistic strange metal description from our model.}

\begin{document} 
\maketitle
\flushbottom

\section{Introduction}\label{sec:intro}
The AdS/CMT correspondence is a string theory-based framework that emerged as a promising tool for studying strongly coupled condensed matter systems. One of the central and original motivations of the AdS/CMT is aiming to reveal the mysterious aspects of strange metal \cite{Phillips:2022nxs} and the related high-$T_c$ superconductivity in cuprates. Remarkably, the linear-in-T resistivity without saturation at low temperatures and high temperatures is irreconcilable with conventional Fermi liquid, indicating the breakdown of the quasiparticle picture. As particles and locality are typically related, the novel behaviors of strange metal raise the distinct possibility that its clear explanations must abandon the basic building blocks of quantum theory. Thus the AdS/CMT correspondence is a natural tool to make progress on this problem. 

Various efforts in constructing the non-Fermi liquid from holography have been performed \cite{Charmousis:2010zz, Iizuka:2011hg,Davison:2013txa,Jeong:2018tua}. 
Among all types of holographic models,  the Gubser-Rocha model \cite{Gubser:2009qt} with momentum relaxation \cite{Davison:2013txa}  is well-known for its linear-in-T resistivity. However, linear-in-T resistivity is only one puzzle among many peculiar transport anomalies of strange metal.  For a metal to be named strange, the T-linearity must extend far beyond the typical bounds associated with phonon-mediated resistivity. Besides T-linear resistivity, strange metal also exhibits linear-in-B magnetoresistivity at high fields and a quadratic $T^2$-dependence of the inverse Hall angle.  Recently reported measurements of the low-temperature normal state magnetoresistance of $\rm{La_{2-x}Ce_{x}CuO_4}$ \cite{Sarkar2018CorrelationBS} exhibit a linear-in-B behavior, obeying a scaling relation with applied field and temperature. The Hall coefficient $R_H$ of strange metal demonstrates its temperature and field dependence, which is incompatible with that expected in conventional metals.  In a Fermi liquid with only one relaxation rate, the Hall coefficient should be nearly T-independent.  Moreover, the observation of a large Nernst response in the strange-metal state in 2D superconductor $2\text{M}-\text{WS}_{2}$ sharpens the difficulties of strange metal problems in condensed matter physics \cite{Yang2023AnomalousEO}.

In this paper, we mainly study the effect of magnetic fields on holographic transports by developing an analytical Gubser-Rocha black hole solution with magnetic fields. We first review the scaling analyses of magneto-thermoelectric conductivities in Einstein-Maxwell-Dilaton-Axion (EMD-Axion) theory by exploring the idea of two relaxation time scales first used by P. Anderson \cite{Anderson91} to explain the Fermi-liquid behavior of the Hall angle. Later it was elaborated in a holographic model by Donos and Blake \cite{Blake:2014yla}. Unfortunately, it was shown in \cite{Amoretti:2016cad} that the model considered by Donos and Blake cannot reproduce the scalings of the strange metal. The standard classes of EMD-Axion models thus need to be modified to account for the IR physics of strange metals. As a holographic model, the magnetic Gubser-Rocha black hole solution can show us unconventional metal properties that might be universal. 

This paper is organized as follows. In section 2, we perform a qualitative analysis of the magneto-thermoelectric transport in EMD model and hydrodynamic theory. In section 3, we consider an extension of the Gubser-Rocha model with momentum relaxation. From this model, a new black hole solution is found. We would analyze the thermodynamic properties of this black hole. In section 4, we compute the longitudinal electric resistivity, Hall angle and the Nernst signal. At the end, conclusions and discussions will be presented in section 5.

\section{Scaling analyses of Magneto-thermoelectric conductivities}
In this section, we will first briefly review the results in \cite{Amoretti:2016cad} why a wide class of EMD models fails to reproduce the scaling behaviour of cuprates. Then, we will assume there are two time scales and study  Magno-thermoelectric transports in hydrodynamic theory. We will try to understand how the Nernst signal is enhanced in non-Fermi liquid and why magneto-resistivity and Hall coefficients depend on the magnetic field from a holographic point of view.

The general action for a class of EMD models is given as follows 
\begin{equation}
S=\int d^4 x \sqrt{-g}\bigg(R-\frac{1}{2}(\partial \phi)^2-\frac{1}{4}Z(\phi)F^2+\frac{1}{2}Y(\phi)\sum^{2}_{i=1}\partial\psi^2_{i}+V(\phi)\bigg),
\end{equation}
The black hole solution is considered to take the following form
\begin{equation}
ds^2=-U(r)dt^2+U^{-1}(r)dr^2+W(r) (dx^2+dy^2). 
\end{equation}
We would like to study the effect of the magnetic field on several transport coefficients, we take the gauge field as
\be
A=A_0(r) \,\dd t -B\, y\,\dd x,
\ee
and the ansatz for the axions is taken to be
\be
\psi_1=kx,\quad \psi_2=ky.
\ee
The disorder parameter $k$ controls the strength of momentum dissipation. Near the black hole horizon, the gauge field $A_0 \sim  (r-r_0)$ vanishes and $U\sim 4\pi T(r-r_0)$. Note that $r_0$ denotes the location of event horizon. At the asymptotic boundary, the gauge field expanded as $A_0 \sim \mu-\frac{n}{r}$ with conserved charge density $n:=W Z A_0'$. 
The general analytic solution for this model was obtained in \cite{Blake:2015ina,Kim:2015wba}. Here we reproduce the expression for the longitudinal resistivity and the Hall angle
\begin{align}
    \begin{split}
        \rho_{xx}&=\frac{k^2 W Y(n^2+k^2 W Y Z+B^2 Z^2)}{((n^2 + Z W Y(\phi) k^2)^2 +
  Z^2 n^2 B^2)}\bigg|_{r_0},\\
        \cot\theta_H&=\frac{k^2 W Y\left(Z\left(B^2 Z+k^2 W Y\right)+n^2\right)}{B n\left(Z\left(B^2 Z+2 k^2 W Y\right)+n^2\right)}\bigg|_{r_0},
    \end{split}
\end{align}
At $B=0$, the DC conductivity obeys the inverse Matthiessen rule
\be
\sigma_{DC}=\sigma_{ccs}+\sigma_{diss}=Z+\frac{n^2}{k^2 W Y}\bigg|_{r_0}.
\ee
As in the two-time scale model, the first term in the conductivity is contributed by the particle-hole pair production, which may refer to ``charge-conjugation symmetric'" term, while the second term is related to momentum dissipation mechanism. Note the physical picture of the particle hole pairs moving induced current is suggestive in the holographic model.   
The Nernst signal is obtained as 
\begin{align}\label{eNdef}
\begin{split}
e_N=(\sigma^{-1}\cdot\alpha)_{xy}=&\frac{(4 \pi Z^2 W^2 B Y(\phi) k^2)}{((n^2 + Z W Y(\phi) k^2)^2 +
  Z^2 n^2 B^2)}\bigg|_{r_0}\\
  =&\frac{1}{k^2 Y}\bigg(\frac{1}{4\pi B }+\frac{B \sigma_{diss}^2 }{4 \pi n^2}+\frac{ \sigma_ {diss}}{2 B \pi \sigma_{ccs}}+\frac{ \sigma _{diss}^2}{4\pi B  \sigma_ {ccs}^2}\bigg)^{-1}\bigg|_{r_0}.
\end{split}
\end{align}
It turns out that the scaling of the transport coefficients of class I model in \cite{Gouteraux:2014hca} are of the Fermi-liquid type, say
\be
\rho_{xx}\propto\cot\theta_H\propto T^\lambda.
\ee
The Nernst signal is of the form
\be
e_N\sim\frac{ T^\kappa}{B^2 T^{2\lambda}+c}.
\ee
If we assume the resistivity is linear in T, i.e. $\lambda=1$, then we must have the condition $0<\kappa<2$ for the Nernst signal to be bell-shaped. From the above analysis, it seems that the standard EMD theory could be successful in reproducing linear-T behavior of the resistivity and bell-shaped curve of the Nernst signal, but fails disastrously in reproducing Fermi-liquid-like Hall angle at the same time.

With the limitations of EMD models, we now turn to hydrodynamic theory specifically studied by \cite{Baggioli:2022pyb} and discuss its implications for models with finite magnetic fields. We show the related transport coefficients obtained in \cite{Amoretti:2019buu}
\begin{align}
\begin{split}
\rho_{x x}&=\frac{1}{\sigma_{ccs}+\sigma_{diss}}+\mathcal{O}\left(B^2\right), \\
 \cot \theta_H&=\frac{n}{B \sigma_{diss}} \frac{1+\frac{\sigma_{ccs}}{\sigma_{diss}}}{1+2 \frac{\sigma_{ccs}}{\sigma_{diss}}}+\mathcal{O}(B), \\
 e_N&=\frac{B \sigma_{ccs} \sigma_{diss}}{n^2\left(\sigma_{ccs}+\sigma_{diss}\right)^2}\left(s \sigma_{ccs}-n \alpha_0\right)+\mathcal{O}\left(B^3\right).
\end{split}
\end{align}
In order to mimic the transport properties of strange metals observed in experiments, we assume the scaling behaviour of related functions to be
\be
\sigma_{ccs}= \frac{1}{T},\quad \sigma_{diss}\sim \frac{1}{T^\alpha},\quad n\sim T^\theta.
\ee
Here $-\alpha$ and $\theta$ are the scaling dimensions of $\sigma_{diss}$ and charge density $n$.

The scaling dimension of the Hall angle is related to $\alpha$ and $\theta$:
\be
\cot\theta_H\sim\frac{n}{B \sigma_{diss}}\sim T^{\beta}\sim T^{\alpha+\theta}
\ee
with $1<\beta\leq 2$.  We mainly consider two options \footnote{Due to its unfavourable properties, we will not discuss Option 1 of \cite{Ahn:2023ciq}.} of the choice of $\{\alpha,\theta\}$ and discuss how these choices would affect the magnetothermoelectric transport. 

$\bf{Option\,I}$: $\alpha=2,\,\theta=0$. Notice that the temperature scaling of $\sigma_{ccs}$ and $\sigma_{diss}$ scales differently in this option. The idea of a two-scale structure dates back to the seminal work of Anderson \cite{Anderson91}. This choice of the scaling is also in accordance with the scaling analysis for Lifshitz-Hyperscaling critical points \cite{Hartnoll:2015sea,Karch:2015zqd}.

$\bf{Option\,II}$: $0<\alpha\leq1,\,0<\theta\leq 1$. The charge density is believed to be temperature independent. But recent experiments \cite{Amoretti:2019buu} show that it may not always be the case. It turns out that it's already been discussed back in 1980s \cite{Cheong87,Stormer88} that the linear-T behaviour of the charge density could account for the Fermi-liquid behaviour of the Hall angle. Therefore we will discuss this scenario as well.

Under Option I, the temperature dependence of $\rho_{xx}$ for weak magnetic field is of the form
\be
\rho_{xx}\propto \frac{T^2}{W+T},
\ee
which is linear for $T\gg W$ while under Option II, we have
\be
\rho_{xx}\propto \frac{T^{\alpha+1}}{W T+T^\alpha},
\ee
which is linear in T for low temperature. For high temperatures, the holographic description is not valid and the T-linear behaviour mainly comes from phonon scattering for temperature higher than the Bloch-Gr$\ddot{\text{u}}$neisen temperature \cite{Hwang19,Hartnoll:2021ydi}. 

The presence of the charge-conjugation symmetric term $\sigma_{ccs}$ can in principle enhance the detected Nernst signal. In the ccs dominated regime $\sigma_{ccs}\gg\sigma_{diss}$, the Nernst signal behaves like
\be
e_{Ncon}\sim\frac{B}{n^2}\sigma_{diss}(s+\frac{n\mu}{T}) ,
\ee
If we further assume $s\sim T^\gamma$ with $1\leq \gamma< 2$, then for Option I 
\be\label{eNcon1}
e_{Ncon}\sim T^{\gamma-2}
\ee
Notice that the $\sigma_{ccs}$-dominated regime corresponds to the high-temperature regime for Option I but corresponds to the low-temperature regime for Option II. For Option II, we have
\be\label{eNcon2}
e_{Ncon}\sim T^{-1-\theta-\alpha}=T^{-1-\beta},
\ee
which is in direct tension with experimental observations.
In dissipation dominated regime, the Nernst signal is dominated by the contribution
\be
e_{Ndiss}\sim\frac{B}{n^2}\frac{\sigma_{ccs}^2}{\sigma_{diss}}(s+\frac{n\mu}{T}),
\ee

which scales as 
\be\label{eNdiss}
e_{ Ndiss } \sim \begin{cases}T^{-1} & \text { for Option I } \\ T^{\alpha+\gamma-2\theta-2} & \text { for Option II }\end{cases}.
\ee
From (\ref{eNcon1}), (\ref{eNcon2}) and (\ref{eNdiss}), the asymptotic behaviour of the Nerst signal does not behave like the non-Fermi liquid either for Option I or Option II. Since the Nernst signal is only approximated to the first order of the magnetic field in this hydrodynamic approach, to fully analyze it we may need to find an explicit analytical formula or use numerical methods. However, since the magnetic field $B$ is coupled to the charge-conjugation symmetric term, the Nernst signal can be strongly enhanced due to this coupling. This in turn may explain the observed enhancement of the Nernst signal.

\section{An S-dual holographic EMD-Axion model}
With the discussions of the above section in mind, we need to point out that the study of the four classes of EMD theories discussed in \cite{Gouteraux:2014hca} is not the end story. In recent years there have been certain attempts \cite{Cremonini:2017qwq,Blauvelt:2017koq,Baggioli:2016oju,Taylor:2014tka,Baggioli:2014roa,Cremonini:2018kla} of exploring other EMD-like theories and apply them to certain condensed matter models by modifying the Maxwell sector and the axion sector of the standard EMD model. With the limitations of the Gubser-Rocha model pointed out in \cite{Ahn:2023ciq}, it's necessary to look for a certain extension of the model. Here we will present the derivation of a dyonic Gubser-Rocha black hole solution. The physical properties of this model will be studied in detail. 
\subsection{A new dyonic AdS black hole solution}
The Gubser-Rocha model with momentum relaxation is a special type of EMD-Axion model which has been studied extensively in the AdS/CMT literature \cite{Davison:2013txa,Jeong:2018tua,Zhou:2015qui,Kim:2017dgz,Cremonini:2016bqw,Jeong:2021wiu}. The analytical solution of this model has been found for zero magnetic field. However, for non-zero magnetic field, we need to rely on numerical methods and no analytical result has been found to this date. Therefore with this in mind, we consider an extended version of this model by introducing an extra axion which couples to the dilaton and the 1-form field. The action reads
\begin{align} \label{action1}
\begin{split}
S & =  S_1 + S_2  = \frac{1}{2\kappa} \int \dd^4x\sqrt{-g}\left( \mathcal{L}_1 + \mathcal{L}_2   \right) \,, \\ 
\mathcal{L}_1 &= R \,-\,\frac{3}{2}\frac{\partial_\mu\tau\partial^\mu\bar{\tau}}{(\text{Im}\tau)^2} \,-\, \frac{1}{4} e^{-\phi}\,F^2 \,+\, \frac{1}{4} \chi\,F\,\tilde{F}+\, \frac{3}{L^2}\frac{\tau\bar{\tau}+1}{\text{Im}\tau} \,, \\
\mathcal{L}_2 &= -\frac{1}{2}\sum_{I=1}^{2}(\partial \psi_{I})^2 \,
\end{split}
\end{align}
where $\tau\equiv\tau_1+i\,\tau_2\equiv \chi+ i\,e^{-\phi}$ is the axio-dilaton, $F_{\mu\nu}$ is the field strength for the vector field $A_\mu$, $\tilde{F}^{\mu\nu}=\frac{1}{2\sqrt{-g}}\epsilon^{\mu\nu\rho\sigma}F_{\rho\sigma}$ is the dual field strength with the convention $\epsilon^{txyr}=1$ and $\psi_1$ and $\psi_2$ are spatially dependent axions that belong to separate sectors. By expanding $\tau$ in $\mathcal{L}_1$,
\begin{align}
\begin{split}
\mathcal{L}_1 =& R \,-\,\frac{3}{2}(\partial\phi)^2 \,-\,\frac{3}{2}e^{2\phi}\,(\partial\chi)^2\,- \frac{1}{4} e^{-\phi}\,F^2 \,\\
+&\, \frac{1}{4} \chi\,F\,\tilde{F}+\, \frac{1}{L^2}(6\cosh\phi+3\chi^2\,e^\phi),
\end{split}
\end{align}
we see that the original Gubser-Rocha action \cite{Gubser:2009qt} is contained in this model. One can check the equations of motion are invariant under the transform
\begin{align}\label{dt}
\begin{split}
\tau&\rightarrow\tau'=-\frac{1}{\tau},\\
F_{\mu\nu}&\ra F'_{\mu\nu}=\tau_1 F_{\mu\nu}+\tau_2 \tilde{F}_{\mu\nu},\\
\tilde{F}_{\mu\nu}&\ra \tilde{F}'_{\mu\nu}=\tau_1 \tilde{F}_{\mu\nu}-\tau_2 F_{\mu\nu},
\end{split}
\end{align}
therefore the first action $S_1$ can be viewed as an "S-dual completion" of the original Gubser-Rocha model. The kinetic term is invariant under the full $\text{SL}(2,\mathbb{R})$ transformation \cite{Sen:1994fa} while the potential term breaks the "T-symmetry"
$\tau\ra\tau+a$. One may restore the full $\text{SL}(2,\mathbb{R})$ symmetry by taking the $L\to\infty$ limit or simply demanding the equivalence relation $\chi\sim\chi+a$. Nevertheless, it suffices to only consider S-duality to obtain a set of solvable equations of motion. For holographic models that do enjoy $\text{SL}(2,\mathbb{R})$ and $\text{SL}(2,\mathbb{Z})$ symmetry, one can consult \cite{Goldstein:2010aw,Fujita:2012fp}. Notice that the original Gubser-Rocha model can be obtained from dimensional reduction of higher-dimensional supergravity theories, whether or not our model is embeddable into higher dimensional string models and if embeddable, how to embed is beyond the scope of this paper. From our limited attempts, at least any naive reduction from gauged supergravity doesn't seem plausible. We encourage curious readers to pursue this direction further. 
Now we are ready to solve the equations of motion by considering the following ansatz
\begin{align} \label{ansatz}
\begin{split}
&\dd s^2  = - f(r) \dd t^2+\frac{1}{f(r)}\dd r^2+g(r)(\dd x^2+\dd y^2), \\ 
&A=A_0(r) \,\dd t -B\, y\,\dd x,\\
&\psi_{1}= k \, x \,, \quad \psi_{2}= k \, y \,, \,
\end{split}
\end{align}
and the axio-dilaton $\tau$ as a function solely of r. Here we assume the other two axions $\psi_1$ and $\psi_2$ share the same momentum relaxation. Anisotropic holographic models has been explored in \cite{Taylor:2014tka,Ge:2014aza,Cheng:2014tya,Andrade:2013gsa}, which we will not discuss this case here. The only non-vanishing component of the equations of motion for the vector field
\be
D_\mu(\tau_2\,F_{\mu\nu}-\tau_1\tilde{F}_{\mu\nu})=0
\ee
takes the following form
\be\label{M2}
A_0^{\prime} g^{\prime}(r)+g(r)\left(A_0^{\prime \prime}-A_0{ }^{\prime} \phi^{\prime}\right)-B e^{\phi} \chi^{\prime}=0.
\ee
By solving equation (\ref{M2}), we can express one function in terms of the other three explicitly, namely
\begin{equation}\label{A0sol}
A_0=\int \dd r \frac{e^{\phi}(C+B \chi)}{g(r)}.
\end{equation}
The rest equations are quite difficult to solve directly, so we further assume the form of $A_0$ and $g(r)$ are unchanged from the usual Gubser-Rocha model:
\be\label{ansatz2}
A_0=\mu\bigg(1-\frac{r_0+\rho}{r+\rho}\bigg),\quad g(r)=\sqrt{r(r+\rho)^3}.
\ee
From this we rewrite the 1-form as
\be\label{oneform}
A=\mu\bigg(1-\frac{r_0+\rho}{r+\rho}\bigg)\,\dd t -P(r_0+\rho)\, y\,\dd x,
\ee
where we denote $B\equiv P(r_0+\rho)$ for later purposes. By multiplying the rr-component of the Einstein equation by $f(r)^2$ and add this to the tt-component of the Einstein equation, we obtain a rather simple equation involving only $g(r)$ and $\tau$:
\begin{align}\label{combE}
\begin{split}
&g^{\prime}(r)^2-g(r)\left(2 g^{\prime \prime}(r)+3 g(r)(\text{Im}\tau)^{-2}\tau'\bar{\tau}'\right)\\
=&g^{\prime}(r)^2-g(r)\left(2 g^{\prime \prime}(r)+3 g(r)\left(\phi'^2+e^{2 \phi} \chi'^2\right)\right)=0.
\end{split}
\end{align}
Plug (\ref{ansatz2}) into (\ref{M2}) and (\ref{combE}), one can get
\be
\chi=C_1+\frac{C_2 \mu\, r\,\left(\mu^2+P^2\right)}{P^3(\rho+r)+\mu^2 P r},\quad
\phi=\log \left(r+\frac{P^2 \rho}{\mu^2+P^2}\right)-\frac{1}{2} \log \left(C_2^2 r(\rho+r)\right),
\ee
where the constants can be determined by taking the $P\to0$ limit and compare the result with the purely electric case \cite{Gubser:2009qt}. After fixing the constants, we obtain
\begin{equation}\label{tausol}
\tau=\frac{-\mu\, P\, \rho+i\left(\mu^2+P^2\right) \sqrt{r(\rho+r)}}{(\mu^2+P^2)r+P^2\rho}.
\end{equation}
Finally, we find that $f(r)$ which satisfies all the rest equations is given by
\be\label{fsol}
f(r)= \sqrt{r(r+\rho)^3}\left(\frac{1}{L^2} - \frac{k^2}{2(r+\rho)^2} -\frac{(\mu^2+P^2)\,(r_0+\rho)^2}{3\rho(r+\rho)^3}  \right).
\ee
Since the form of the metric doesn't change from the purely electric case, the near-horizon geometry is still conformal to $\text{AdS}_2\times \mathbb{R}^2$ \cite{Andrade:2013gsa,Gouteraux:2014hca}. 
Acting the duality transform (\ref{dt}) on (\ref{oneform}) and (\ref{tausol}), which is essentially 
\be\label{paratransform}
\left(\begin{array}{c}
\mu \\
P
\end{array}\right) \rightarrow\left(\begin{array}{cc}
0 & 1 \\
-1 & 0
\end{array}\right)\left(\begin{array}{l}
\mu \\
P
\end{array}\right)=\left(\begin{array}{c}
P \\
-\mu
\end{array}\right),
\ee
we would get a dual solution. Taking $\mu=0$ and $P=0$  corresponds to the purely magnetic and purely electric case respectively, where the purely electric case simply corresponds to the usual Gubser-Rocha model.
\subsection{Thermodynamics}
\begin{figure}[t]
\begin{center}
  \begin{minipage}[b]{0.45\linewidth}
    \centering
    \includegraphics[height=3.9cm]{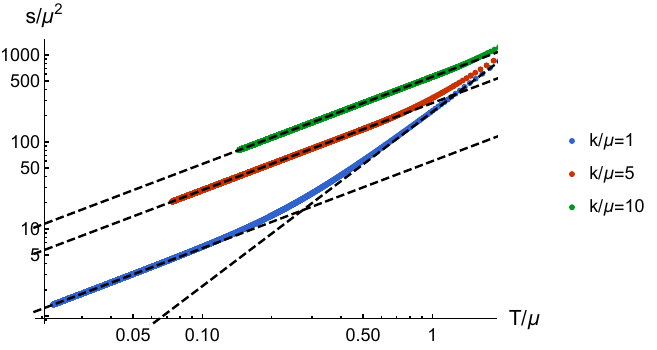}
  \end{minipage} \hspace{0.5cm}
  \begin{minipage}[b]{0.45\linewidth}
    \centering
    \includegraphics[height=3.9cm]{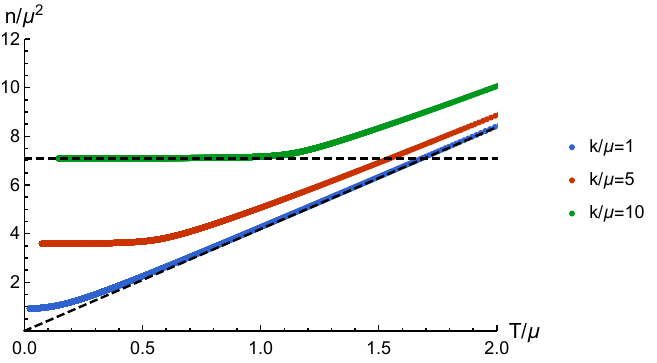}
  \end{minipage} 
\end{center}
\caption{Left:temperature dependence of entropy density for various momentum relaxation at $\bar{B}=0.1$. Right:temperature dependence of charge density for various momentum relaxation at $\bar{B}=0.1$}
  \label{sbnbTb}
\end{figure}

From the metric, the temperature and entropy density of the black hole/boundary theory can be obtained
\be\label{Tands}
T=\frac{f'(r_0)}{4\pi},\qquad s=4\pi g(r_0),
\ee 
where $r_0$ is the horizon value where $f(r)=0$. Using (\ref{fsol}), we have
\be
\frac{1}{L^2} =\frac{\mu^2}{3\rho(r_0+\rho)}+ \frac{k^2}{2(r_0+\rho)^2}  +\frac{B^2}{3\rho(r_0+\rho)^3} .
\ee
 Due to the presence of the topological term $\chi \mathrm{F}\wedge \mathrm{F}$, the conserved quantity that should be interpreted as charge density gets modified to
\be\label{chargedensity}
n=g(r)\,e^{-\phi}A_0'-B\,\chi,
\ee
which is still given by $\mu(r_0+\rho)$ when evaluated at the horizon. Fixing $\bar{k}=k/\mu$ and $\bar{B}=B/\mu^2\neq0$ and set $L=1$, the dimensionless chemical potential can be solved as
\begin{equation}\label{chempo}
\tilde{\mu}:=\frac{\mu}{r_0}=\frac{1}{2 \bar{B}} \sqrt{(1+\tilde{\rho})\left(\sqrt{48 \bar{B}^2 \tilde{\rho}(1+\tilde{\rho})+\left(2+\left(2+3 \overline{\mathrm{k}}^2\right) \tilde{\rho}\right)^2}-2-\left(2+3 \overline{\mathrm{k}}^2\right) \tilde{\rho}\right)}
\end{equation}
where $\tilde{\rho}:=\rho/r_0$. From (\ref{Tands}), (\ref{chargedensity}) and (\ref{chempo}), we get the rescaled temperature
\begin{equation}
\begin{aligned}
\bar{T}:=\frac{T}{\mu}=& \frac{3 \bar{B}}{2 \pi \sqrt{\sqrt{4+\tilde{\rho}\left(48 \overline{\mathrm{B}}^2(1+\tilde{\rho})+\left(2+3 \overline{\mathrm{k}}^2\right)\left(4+\left(2+3 \overline{\mathrm{k}}^2\right) \tilde{\rho}\right)\right)}-2-\left(2+3 \overline{\mathrm{k}}^2\right) \tilde{\rho}}} \\
 -&\frac{\overline{\mathrm{k}}^2 \sqrt{\sqrt{4+\tilde{\rho}\left(48 \overline{\mathrm{B}}^2(1+\tilde{\rho})+\left(2+3 \overline{\mathrm{k}}^2\right)\left(4+\left(2+3 \overline{\mathrm{k}}^2\right) \tilde{\rho}\right)\right)}-2-\left(2+3 \overline{\mathrm{k}}^2\right) \tilde{\rho}}}{16 \pi \overline{\mathrm{B}}(1+\tilde{\rho})}
\end{aligned}
\end{equation}
and dimensionless entropy density and charge density
\begin{equation}\label{sn}
\begin{aligned}
\bar{s}&:=\frac{s}{\mu^2}= \frac{16 \pi \bar{B}^2 \sqrt{1+\tilde{\rho}}}{\sqrt{4+4\left(2+12 \bar{B}^2+3 \bar{k}^2\right) \tilde{\rho}+\left(48 \bar{B}^2+\left(2+3 \bar{k}^2\right)^2\right) \tilde{\rho}^2}-2-\left(2+3 \bar{k}^2\right) \tilde{\rho}},\\
\bar{n}&:= \frac{n}{\mu^2}=\frac{2 \bar{B} \sqrt{\tilde{\rho}+1}}{\sqrt{\sqrt{48 \bar{B}^2 \tilde{\rho}(\tilde{\rho}+1)+\left(\left(3 \bar{k}^2+2\right) \tilde{\rho}+2\right)^2}-\left(3 \bar{k}^2+2\right) \tilde{\rho}-2}}.
\end{aligned}
\end{equation}
We can plot the entropy density and charge density as a function of rescaled temperature, see Fig.\ref{sbnbTb}. The asymptotic scaling behaviour of these functions agree with the numerical results for undualized model \cite{Ahn:2023ciq}. For $\bar{T}\ll 1$, $\bar{s}\sim\bar{T}$ and $\bar{n}\sim\bar{T}^0$ while for $\bar{T}\gg 1$, $\bar{s}\sim\bar{T}^2$ and $\bar{n}\sim\bar{T}$. From the expression of (\ref{sn}), we see that the charge density is correlated with the magnetic field. This phenomenon also occurs in the context of AdS/BCFT \cite{Fujita:2012fp,Melnikov:2012tb} and in \cite{Jeong:2023hrb}\footnote{We are grateful to Hyun-Sik Jeong for introducing these works.}.

\section{Thermo-electric transports}
\begin{figure}[t]
\begin{center}
    \centering
    \includegraphics[height=4.6cm]{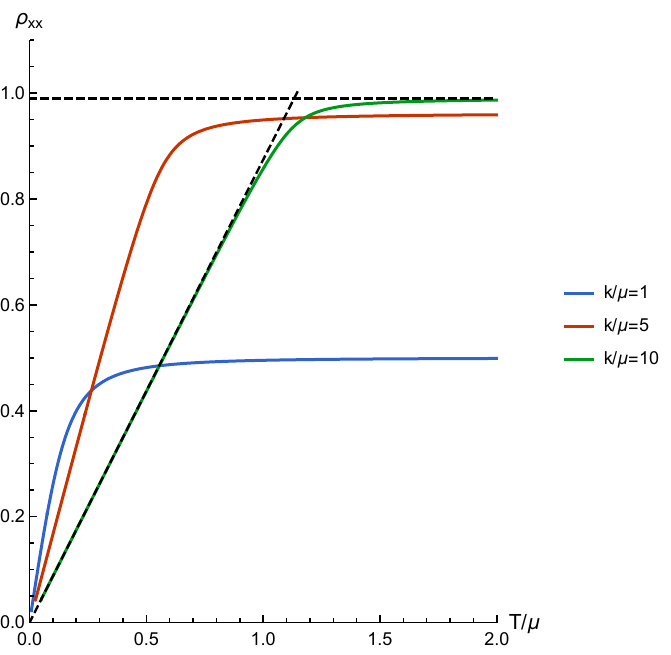}
    \includegraphics[height=4.6cm]{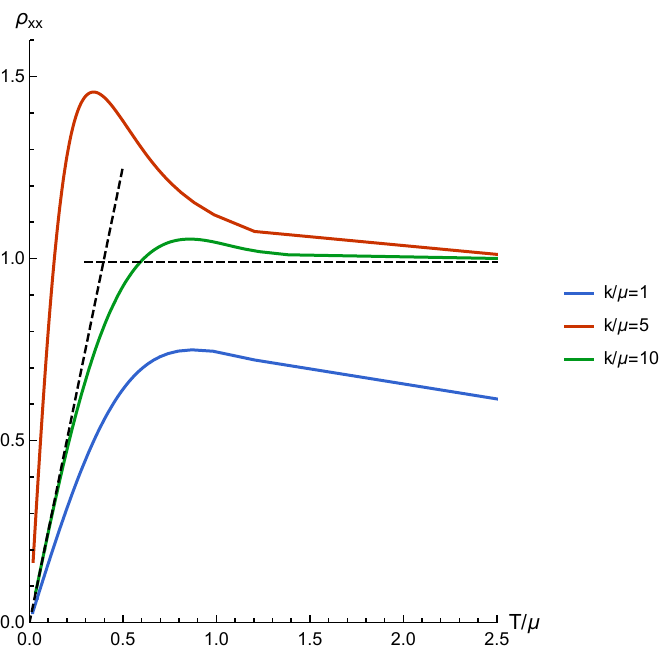}
    \includegraphics[height=4.6cm]{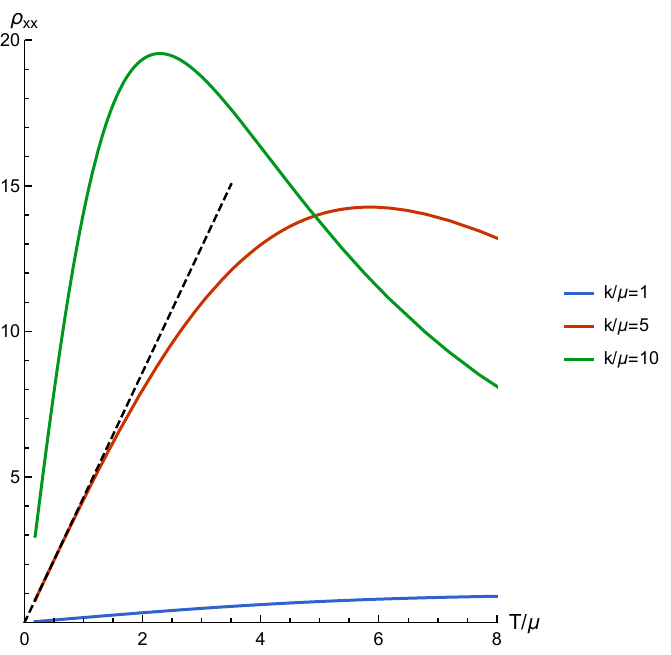}
 \end{center}
\caption{Left:temperature dependence of electric resistivity for various momentum relaxation at $\bar{B}=0.1$. Middle:temperature dependence of electric resistivity for various momentum relaxation at $\bar{B}=10$. Right:temperature dependence of the electric resistivity for various momentum relaxation at $\bar{B}=1000.$}
  \label{fig2}
\end{figure}

\begin{figure}[t]
\begin{center}
 
    \centering
    \includegraphics[height=6cm]{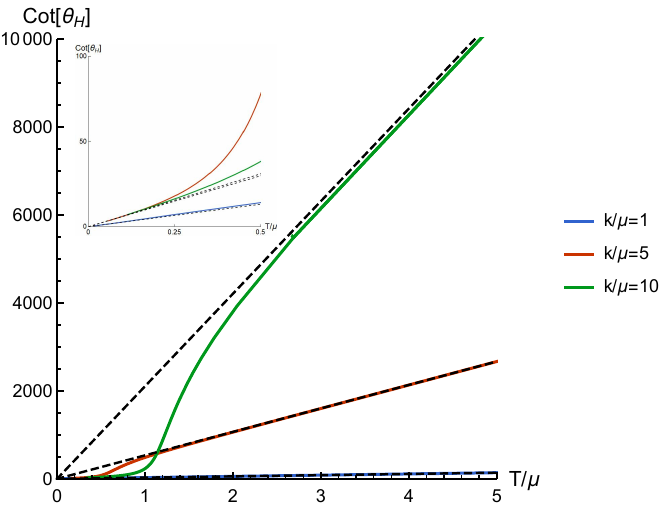}
  
\end{center}
\caption{Temperature dependence of the Hall angle for various momentum relaxation at $\bar{B}=0.1$}
  \label{fig3}
\end{figure}

\begin{figure}[t]
\begin{center}
  \begin{minipage}[b]{0.45\linewidth}
    \centering
    \includegraphics[height=5.2cm]{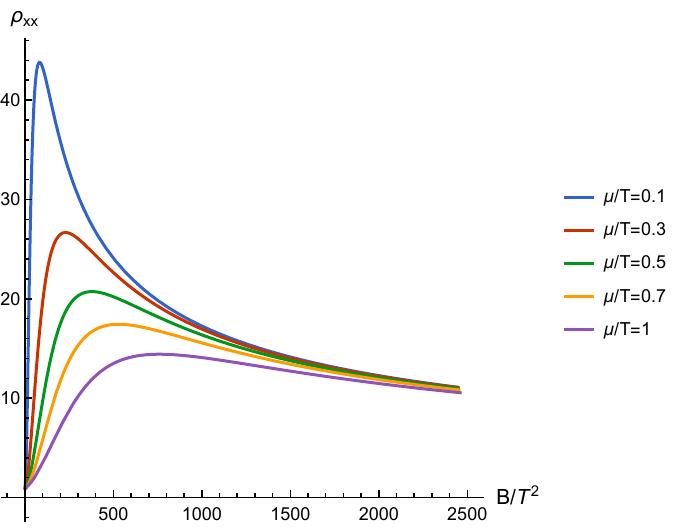}
  \end{minipage} \hspace{0.2cm}
  \begin{minipage}[b]{0.45\linewidth}
    \centering
    \includegraphics[height=5.2cm]{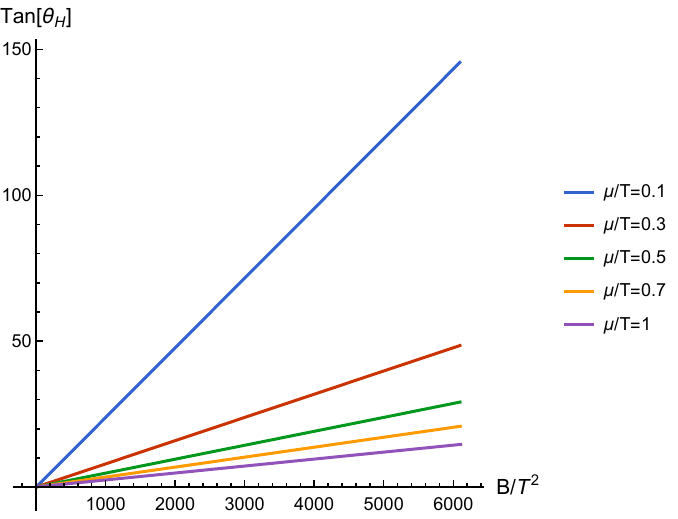}
  \end{minipage} 
\end{center}
\caption{Left: dependence on the magnetic field of electric resistivity for various chemical potentials at $\bar{k}=10$. Right: dependence on the magnetic field of the Hall angle for various chemical potentials at $\bar{k}=10$.}
  \label{fig4}
\end{figure}

We now proceed to calculate the transport properties of the boundary theory using the methods given in \cite{Blake:2013bqa,Blake:2014yla,Donos:2014cya}. We can work with Eddington-Finkelstein like coordinates as in \cite{Donos:2014cya,Chen:2017gsl}, with $v:=t+\int \frac{\dd r}{f}$, the unperturbed metric reads
\be
\dd s^2=-f(r)\dd v^2+2\dd v\,\dd r+g(r)(\dd x^2+\dd y^2).
\ee
Consider the perturbation around the background  to be
\be
\delta g_{v i}=g(r)\,h_{t i}(r)+\mathcal{O}(r-r0),\qquad \delta A_i=-E_i\,v+\mathcal{O}(r-r0), \quad i=x,y,
\ee
the v-x and v-y component of the linearized Einstein equation near the horizon would then give us two linear equations that can be put into matrix form
\begin{align}
\begin{split}
&\left(\begin{array}{cc}
k^2\, g(r_0)\, e^{\phi(r_0)}+B^2 & -B\, g(r_0)\, A_0{ }^{\prime}(r_0) \\
B\, g(r_0)\, A_0{ }^{\prime}(r_0) & k^2\, g(r_0)\, e^{\phi(r_0)}+B^2
\end{array}\right)\left(\begin{array}{l}
h_{tx} \\
h_{ty}
\end{array}\right)\\
=-&\left(\begin{array}{cc}
g(r_0) \,A_0{ }^{\prime}(r_0) & B \\
-B & g(r_0)\, A_0{ }^{\prime}(r_0)
\end{array}\right)\left(\begin{array}{l}
E_x \\
E_y
\end{array}\right).
\end{split}
\end{align}
From this matrix equation, the off-diagonal metric perturbations can be expressed in terms of the applied electric fields. 
The perturbed Maxwell equation would give us two constant currents
\begin{align}
\begin{split}
J_x=\tau_2\,E_x-\tau_1\,E_y-\tau_2\,g(r)A_0'\,h_{tx}-\tau_2 \,B\,h_{ty}\bigg|_{r=r_0},\\
J_y=\tau_1\,E_x+\tau_2\,E_y+\tau_2\,B\,h_{tx}+\tau_2 \,g(r)A_0'\,h_{ty}\bigg|_{r=r_0}.\\
\end{split}
\end{align}
Using the generalised Ohm's law \footnote{For the purpose of this paper it suffices to only discuss $\sigma_{ij}$ and $\alpha_{ij}$. }
\begin{equation}
\left(\begin{array}{c}
\mathrm{J} \\
\mathrm{Q}
\end{array}\right)=\left(\begin{array}{cc}
\sigma & \alpha T \\
\bar{\alpha} T & \bar{\kappa} T
\end{array}\right)\left(\begin{array}{c}
\mathrm{E} \\
-(\nabla T) / T
\end{array}\right),
\end{equation}
together with the expression for the heat current \cite{Hartnoll:2007ih,Donos:2014cya}
\be
Q_i=f^2\left(\frac{\delta g_{ti}}{f}\right)'-A_0\,J_i\bigg|_{r=r_0}=-4\pi\,T\,g(r_0)\,h_{ti}(r_0), \quad i=x,y,
\ee
one can compute the conductivity matrix $\sigma$ and thermo-electric conductivity matrix $\alpha$:
\begin{align}\label{sigmaalpha}
\begin{split}
\sigma_{xx}=\sigma_{yy}=&\frac{k^2\, g\left(r_0\right)\left(B^2+k^2\, g\left(r_0\right) e^{\phi\left(r_0\right)}+g(r_0)^2 A_0^{\prime}\left(r_0\right)^2\right)}{\left(B^2+k^2\,e^{\phi(r_0)} \, g\left(r_0\right)\right)^2+B^2\,g\left(r_0\right)^2 \, A_0^{\prime}\left(r_0\right)^2},\\
\sigma_{xy}=-\sigma_{yx}=&\frac{B \,g(r_0) e^{-\phi(r_ 0)} A_0{ }^{\prime}\left(r_0\right)\left(2 k^2\, g\left(r_0\right) e^{\phi(r_ 0)}+g(r_0)^2\, A_0{ }^{\prime}\left(r_0\right)^2+B^2\right)}{\left(B^2+k^2\,e^{\phi(r_ 0)} \, g\left(r_0\right)\right)^2+B^2\,g\left(r_0\right)^2\, A_0{ }^{\prime}\left(r_0\right)^2}-\chi(r_ 0),\\
\alpha_{xx}=\alpha_{yy}=&\frac{4\pi\,k^2\, g\left(r_0\right)^3 \,e^{\phi(r_0)}\,A_0'(r_0)}{\left(B^2+k^2\,e^{\phi(r_0)} \, g\left(r_0\right)\right)^2+B^2\,g\left(r_0\right)^2 \, A_0^{\prime}\left(r_0\right)^2},\\
\alpha_{xy}=-\alpha_{yx}=&\frac{4\pi\,B\,g(r_0)\,(B^2+k^2\,e^{\phi(r_0)}\,g(r_0)+g(r_0)^2\,A_0'(r_0)^2)}{\left(B^2+k^2\,e^{\phi(r_0)} \, g\left(r_0\right)\right)^2+B^2\,g\left(r_0\right)^2 \, A_0^{\prime}\left(r_0\right)^2}.
\end{split}
\end{align}
Several physical quantities concerning electrical transport can be obtained from (\ref{sigmaalpha}) such as the longitudinal electric resistivity
\begin{align}
\begin{split}
\rho_{xx}&=(\sigma^{-1})_{xx}=\frac{\sigma_{xx}}{\sigma_{xx}^2+\sigma_{xy}^2}\\
&=\frac{k^2\left(P^2+\mu^2\right)\left(k^2+P^2+\mu^2\right) \sqrt{r_0(r_0+\rho)}}{r_0\left(P^2+\mu^2\right)\left(k^4+\left(2 k^2+P^2\right) \mu^2+\mu^4\right)+\mu^2\left(k^2+P^2+\mu^2\right)^2 \rho},
\end{split}
\end{align}
 and the Hall angle
\be
\tan \theta_H\equiv\frac{\sigma_{xy}}{\sigma_{xx}}=\frac{\mu P\left(2 k^2\left(\mu^2+P^2\right)\left(\rho+r_0\right)+k^4 \rho+\left(\mu^2+P^2\right)^2\left(\rho+r_0\right)\right)}{k^2\left(\mu^2+P^2\right) \left(k^2+\mu^2+P^2\right)\sqrt{r_0\left(r_0+\rho\right)}}.
\ee
The temperature dependence of these quantities can be expressed analytically. Since the expression is quite complicated, we plot $\rho_{xx}$ vs $\bar{T}$ and $\cot\theta_H$ vs $\bar{T}$ to examine their scaling behaviours, see Fig.\ref{fig2} and Fig.\ref{fig3}.
We can see that the scaling behaviour of the Hall angle improves over the undualized model. Near zero temperature, $\tan\theta_H\simeq c_1\, \bar{T}$ and for high temperature regime $\tan\theta_H\simeq c_2\,\bar{T}$. Here the proportional constants are in general not equal to each other, i.e. $c_1\neq c_2$. Meanwhile, the asymptotic scaling behaviour of $\rho_{xx}$ for the weak magnetic field agrees with the undualized model. For strong magnetic field and large momentum relaxation, one can extend the linear-T resistivity behaviour into higher temperature. However, our model would be in the insulator phase for high temperature and strong magnetic field. For high temperature regime, one can show explicitly that
\be
\rho_{xx}\to \frac{\bar{k}^2}{1+\bar{k}^2} ,
\ee
 when $\bar{T}\to\infty$ with fixed $\bar{B}$ and $\bar{k}$. In the high temperature regime, the resistivity tends to a constant value, strongly deviates from the T-linear behavior. We remark here that the Hall angle is invariant under S-duality (\ref{paratransform}) up to a negative sign. This can be understood by the fact that the complex combinations of the conductivity $\sigma_{\pm}=\sigma_{xy}\pm i\sigma_{xx}$ which transform in the same way as $\tau$, are the S-dual of the complexified resistivities $\rho_{\pm}=\rho_{xy}\pm i\rho_{xx}$. Though the physical parameters we are interested in are $\mu$ and $B$, the duality we see here is an effective way to make our computation non-perturbative in $B$. In Fig.\ref{fig4}, we plot $\rho_{xx}$ and $\tan\theta_H$ as functions of $B/T^2$. 
The dependence on the magnetic field of $\rho_{xx}$ and $\tan\theta_H$ is shown in Fig.\ref{fig4}. We see that $\tan\theta_H$ shown in the right figure of Fig.\ref{fig4}  is almost perfectly proportional to the magnetic field. 
\subsection{Nernst effect}
The Nernst signal signaling the Nernst effect, which has been displayed in (\ref{eNdef}) can also be defined as in \cite{Kim:2015wba} 
\be
e_N:=(\sigma^{-1}\cdot\alpha)_{xy},
\ee
for our model. It's explicit expression can be calculated, which is given by
\begin{equation}
e_N=\frac{4\pi k^2 P  \left(P^2+\mu^2\right)r_0(r_0+\rho)}{\left(P^2+\mu^2\right)\left(k^4+\left(2 k^2+P^2\right) \mu^2+\mu^4\right)r_0+\mu^2\left(k^2+P^2+\mu^2\right)^2 \rho}.
\end{equation}
\begin{figure}[t]
\begin{center}
  \begin{minipage}[b]{0.45\linewidth}
    \centering
    \includegraphics[height=5.5cm]{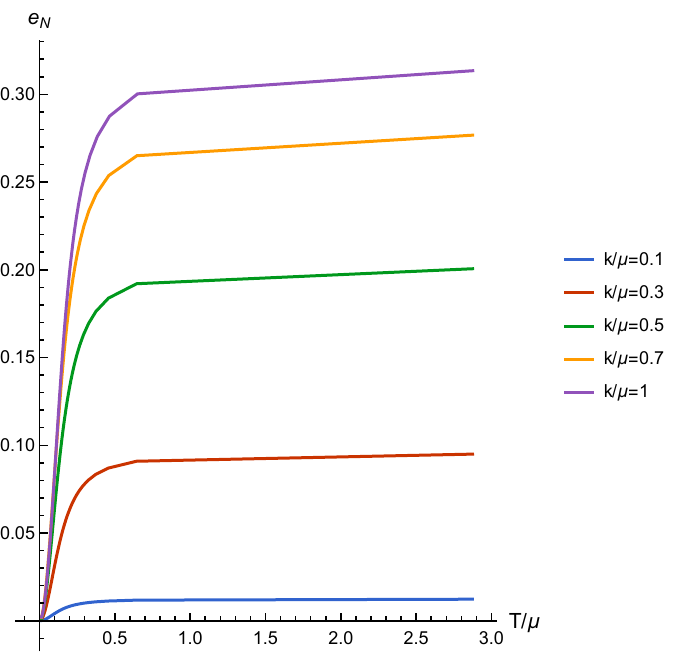}\\
    (a)
  \end{minipage} \hspace{0.2cm}
  \begin{minipage}[b]{0.45\linewidth}
    \centering
    \includegraphics[height=5.5cm]{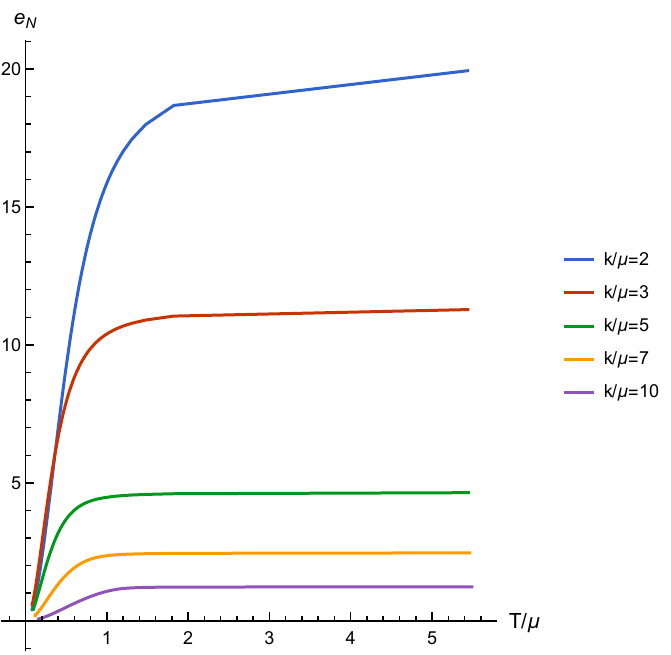}\\
    (b)
  \end{minipage}
  
\end{center}
\caption{Temperature dependence of the Nernst signal for various momentum relaxation at (a) $\bar{B}=0.1$, (b) $\bar{B}=10$.}
  \label{fig5}
\end{figure}

\begin{figure}[t]
\begin{center}
  \begin{minipage}[b]{0.45\linewidth}
    \centering
    \includegraphics[height=5.2cm]{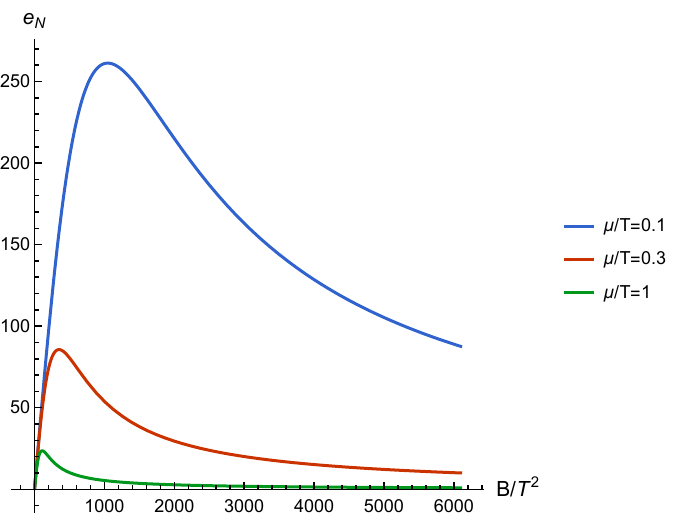}
  \end{minipage} \hspace{0.2cm}
  \begin{minipage}[b]{0.45\linewidth}
    \centering
    \includegraphics[height=5.2cm]{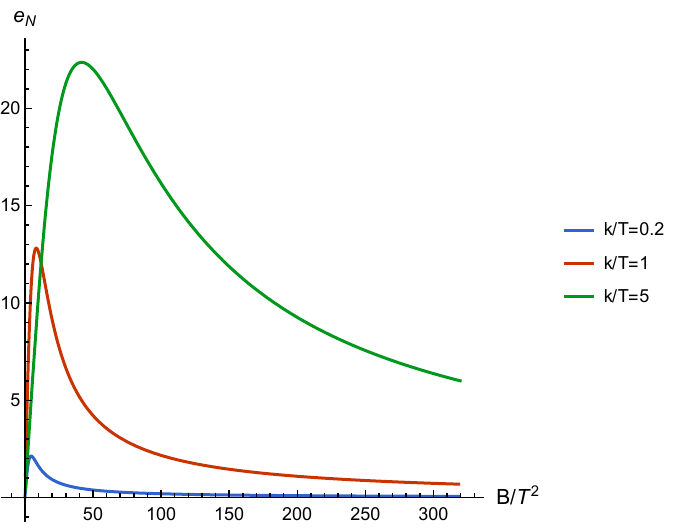}
  \end{minipage} 
\end{center}
\caption{Left: dependence of the Nernst signal on the magnetic field for various chemical potentials for $k/T=5$; Right: dependence of the Nernst signal on the magnetic field for various momentum dissipations for $\mu/T=1$. }
  \label{fig6}
\end{figure}
In Fig. \ref{fig5},  we plot the Nernst signal as a function of $\frac{T}{\mu}$ in the weak and strong magnetic field limit, respectively.  The temperature behavior of the Nernst signal (See Fig. \ref{fig5}) also has this "flattening" phenomenon which can be understood by that the Nernst signal is constructed out of the resistivity matrix and indeed, the asymptotic behavior of the Nernst signal shows
\be\label{asymbofeN}
e_N\to \frac{4\pi\bar{B}\bar{k}^2}{(1+\bar{k}^2)^2},
\ee
as $\bar{T}\to\infty$. Notice that the RHS of (\ref{asymbofeN}) as a function of $\bar{k}$ reaches its maximum for $\bar{k}=1$, therefore we have another peculiar property of our model that the Nernst signal got reduced when we increase the momentum dissipation after it surpasses $\bar{k}=1$, that is to say $e_{N}\sim \frac{4\pi \bar{B}}{\bar{k}^2}$.

As one can see in Fig.\ref{fig6}, the Nernst signal is bell-shaped as a function of $B/T^2$, and we haven't observed any transition to ordinary metal phase for other choices of parameters. The Nernst signal is enhanced if one increases the momentum dissipation while keeping $\mu/T$ fixed. In contrast, the Nernst signal is reduced when one increases the chemical potential while keeping $k/T$ fixed. 
\section{Conclusions and discussions}
Motivated by the limitations of EMD-Axion model approach to strange-metal phenomenology, we have obtained a new dyonic dilatonic asymptotically AdS black hole solution based on an EMD-Axion model which can be viewed as a dyonic version of the Gubser-Rocha model with momentum relaxation. Here requiring S-duality of the action plays an important role in finding the analytical solution. By taking the zero magnetic field limit, the solution reduces to the usual Gubser-Rocha model with momentum relaxation. The AdS/CMT dictionary can help us investigate the longitudinal electrical resistivity, Hall angle and the Nernst signal of the boundary theory. Our calculations show that $\rho_{xx}$ exhibit linear-T resistivity for low temperature. It turns out that with large momentum relaxation and a strong magnetic field, the temperature at which $\rho_{xx}$ starts to deviate from linear-T behaviour is higher than the case with large momentum relaxation but with a weak magnetic field. The Hall angle $\cot\theta_H$ of our model is linear in T for both the low-temperature regime and the high-temperature regime for $k/\mu$ keeping fixed. $\tan\theta_H$ is also almost perfectly proportional to $B/T^2$ with $k/T$ fixed. The Nernst signal is a bell-shaped function with respect to $B/T^2$ for our model even when the momentum relaxation is large. 

\begin{figure}[t]
\begin{center}
 
    \centering
    \includegraphics[height=7cm,width=8.2cm]{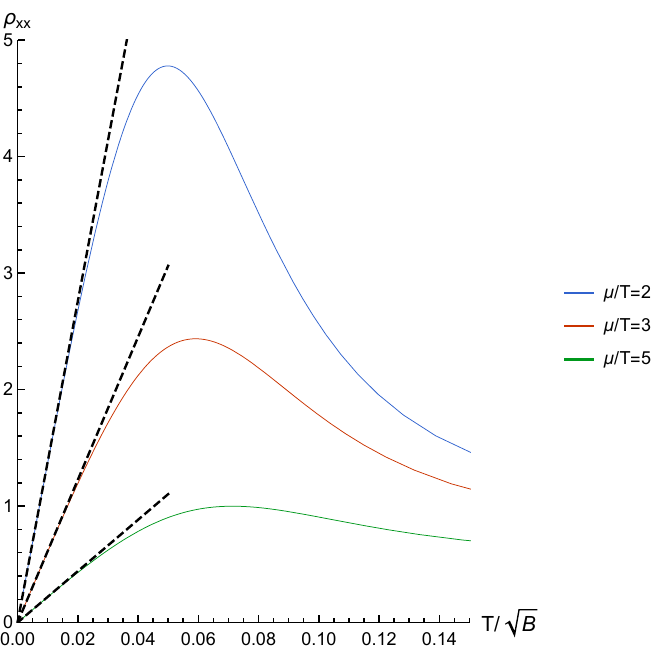}
  
\end{center}
\caption{Temperature dependence of the longitudinal resistivity for various chemical potentials at $k/T=10$}
  \label{fig7}
\end{figure}

Our analytical result seems to resonate with the numerical result done in \cite{Ahn:2023ciq} that the holographic Gubser-Rocha model does not capture all the transport anomalies of strange metals. One possible improvement towards semi-realistic transport properties would be to modify the momentum relaxation $k$ as a function of $r_0$ and $\rho$. As we have already observed in Fig. \ref{fig4}, $\cot\theta_H$ is proportional to $T^2/B$ with $k/T$ and $\mu/T$ being kept constant. And if we plot $\rho_{xx}$ vs $T/\sqrt{B}$ with $k/T$ and $\mu/T$ fixed, see Fig. \ref{fig7}, it also shows linear-T behavior for low temperature. And from Fig. \ref{fig6}, the Nerst signal would be a bell-shaped function in terms of $T/\sqrt{B}$. How to interpret this choice of parametrization in the experimental setup is worth discussing. Another issue we have here is that it's difficult to extend linear-T resistivity into arbitrarily high temperatures, which is inherited from the standard Gubser-Rocha model. Since there's no Debye temperature associated with the holographic model and the linear-T resistivity of the strange metal for high temperature comes from phonons, it's practical to consider the validity of the holographic model only up to a certain scale. In addition, there's no two-scale structure emerging in our model but the observation in \cite{Yang2023AnomalousEO} indicates that there's indeed a two-scale structure associated with the material $2\text{M}-\text{WS}_2$, how to construct a model (not necessarily holographic) describing this type of strange metal would be a more challenging problem. We hope to come back to these issues in later works.

For future directions, one thing that can be explored is to find more black hole solutions based on the action (\ref{action1}). From \cite{Chow:2013gba}, we believe with the presence of S-duality, new spherically symmetric and Kerr-Newman-like AdS black holes can be found. Another thing one can do is to investigate the holographic superconductor \cite{Hartnoll:2008vx,Hartnoll:2008kx} based on this model by including a superconducting phase to the action as in \cite{Jeong:2021wiu}. Furthermore, it would be interesting to compute the quasi-normal modes \cite{Hatsuda:2021gtn}  of this black hole and discuss the corresponding bound on diffusion constant \cite{Blake:2018leo, Jeong:2022luo,Jeong:2021zhz,Jeong:2021zsv,Davison:2016ngz,Blake:2016wvh,Blake:2017qgd,Kim:2017dgz,Hartman:2017hhp,Baggioli:2020ljz,Huh:2021ppg} and the pole-skipping phenomenon \cite{Jeong:2023ynk,Jeong:2021zhz,Grozdanov:2017ajz,Blake:2017ris,Blake:2018leo,Blake:2019otz,Yuan:2023tft}. By including some additional sectors to the dyonic Gubser-Rocha model, one can also discuss its applications to the Kondo condensation effect \cite{Im:2023ffg}.

\acknowledgments
We would like to thank Yongjun Ahn, Shuta Ishigaki, Hong L$\ddot{\text{u}}$, Sang-Jin Sin and Keun-Young Kim for helpful discussions. We are especially grateful to Matteo Baggioli and Hyun-Sik Jeong for giving us valuable comments on the manuscript. This work is partly
supported by NSFC, China (Grant No. 12275166 and Grant. No. 12311540141).

\bibliographystyle{JHEP}

\providecommand{\href}[2]{#2}\begingroup\raggedright\endgroup

\end{document}